\documentclass[3p,times, twocolumn]{elsarticle}

\usepackage{ecrc}
\usepackage{graphicx}
\DeclareGraphicsExtensions{.pdf,.png,.jpg}

\volume{00}

\firstpage{1}

\journalname{Nuclear Physics B Proceedings Supplement}

\runauth{E. Resconi, for the IceCube Collaboration}

\jid{NUPHBP}

\jnltitlelogo{Nuclear Physics B Proceedings Supplement}

\usepackage{amssymb}
\usepackage{lineno}

\usepackage[figuresright]{rotating}

\begin{document}

\begin{frontmatter}

%% Title, authors and addresses

%% use the tnoteref command within \title for footnotes;
%% use the tnotetext command for the associated footnote;
%% use the fnref command within \author or \address for footnotes;
%% use the fntext command for the associated footnote;
%% use the corref command within \author for corresponding author footnotes;
%% use the cortext command for the associated footnote;
%% use the ead command for the email address,
%% and the form \ead[url] for the home page:
%%
%% \title{Title\tnoteref{label1}}
%% \tnotetext[label1]{}
%% \author{Name\corref{cor1}\fnref{label2}}
%% \ead{email address}
%% \ead[url]{home page}
%% \fntext[label2]{}
%% \cortext[cor1]{}
%% \address{Address\fnref{label3}}
%% \fntext[label3]{}

\dochead{}
%% Use \dochead if there is an article header, e.g. \dochead{Short communication}

\title{High Energy Neutrino Astronomy:  IceCube 22 and 40 strings. }

%% use optional labels to link authors explicitly to addresses:
%% \author[label1,label2]{<author name>}
%% \address[label1]{<address>}
%% \address[label2]{<address>}

\author{E. Resconi for the IceCube Collaboration}

\address{MPIK, Saupfercheckweg 1, 69117 Heidelberg (Germany)}

\begin{abstract}
%The construction of the IceCube Neutrino Telescope at the South Pole is over 90\% completed.
%IceCube provides an unprecedented high statistics sample of atmospheric neutrinos spanning five orders of magnitude in energy (100 GeV - 400 TeV), 
%world- wide best sensitivity to galactic and extra-galactic cosmic ray accelerators as well as to the spin-dependent cross section of dark matter with nucleons.
%As of February 2010, 79 strings, 4740 DOMs and 73 IceTop stations comprise IceCube. IceCube includes an inner array (DeepCore) of densely spaced strings,
 %equipped with high quantum efficiency PMTs. The use of DeepCore will allow the lowering of the energy threshold down to about 10 GeV.
 %IceCube enjoys remarkably stable operation: 96-98\% throughout the year, low PMTs dark noise (200-300 Hz), minimal PMTs failure rate 
 %and no evidence of any aging effects.
 In this paper, after a short introduction to the physics of neutrino telescopes, 
 we will report on first performances of the IceCube detector and a selection of preliminary results obtained from data taken while IceCube operated in a partially completed configuration (22 strings and 40 strings). 
We will emphasize new analysis methods recently developed for the study of the Southern Hemisphere as well as for extended regions. 
Based on the long term experience of AMANDA and IceCube, the South Pole ice has proven to be an ideal site for astroparticle physics. 
New ideas and projects about the future beyond IceCube will conclude this presentation.
\end{abstract}

\begin{keyword}
%% keywords here, in the form: keyword \sep keyword
Neutrino astronomy, IceCube
%% MSC codes here, in the form: \MSC code \sep code
%% or \MSC[2008] code \sep code (2000 is the default)
\end{keyword}

\end{frontmatter}

%%
%% Start line numbering here if you want
%%
% \linenumbers

%% main text
\section{Introduction}
\label{Intro}
\begin{center}
\begin{figure}[ht]
\centering
\includegraphics[scale=0.26]{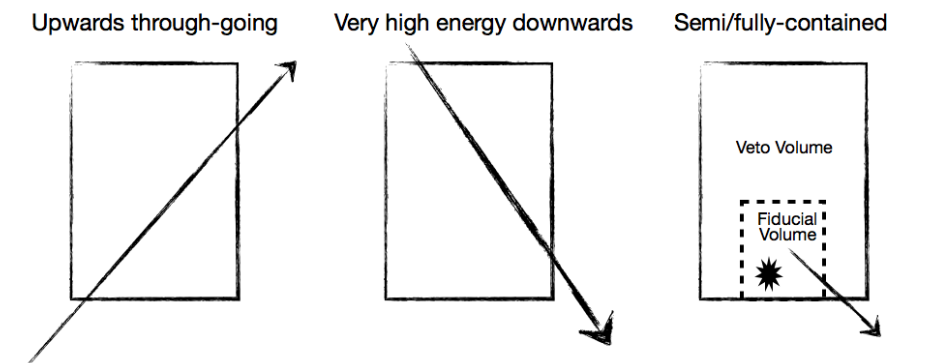}
\caption{Event topologies available in a cubic-kilometer scale neutrino telescope.}
\label{fig:signatures}
\end{figure}
\end{center}

%%%%%%%%%%%%%%Neutrino Astronomy\\
In contrast to photons \cite{JV}, high-energy (HE) neutrinos (E$_\nu>$100~GeV) carry an unambiguous signature for  both acceleration and interaction of  protons in cosmic sites.
The long standing problem of the origin and production mechanisms of cosmic rays \cite{VB} motivates the use of HE neutrinos as probe for a deep investigation of the non-thermal universe \cite{f1}. High energy neutrinos are expected to be produced in the decay of pions, created through proton-proton or proton-photon interactions. 
Candidate cosmic accelerators are both extra-galactic objects like Active Galactic Nuclei (AGN) and Gamma Ray Bursts (GRBs) as well as galactic sources like micro-quasars and supernovae remnants (SNRs). For a review of candidate astronomical sources, we refer to \cite{teresa}. 
\noindent
On the other hand, HE neutrinos are also produced by the annihilation of dark matter particles gravitationally trapped at the center of the Sun, the Earth and our galaxy. For a review about indirect dark matter search with neutrino telescopes we refer to \cite{f2}.
\noindent
Moreover, HE cosmic neutrinos present a unique opportunity to study the interactions of elementary particles at energies comparable and beyond those obtained in current or planned colliders. Lorentz symmetry violation (LV), neutrino decay, monopole, double Sleptons are examples of possible phenomena that can be probed with HE neutrinos. For a description of the phenomenology involved we refer to \cite{LIV} \cite{SUSY}.\\
\noindent
The broad discovery potential of HE neutrino astronomy combined with the weak interaction of neutrinos with matter motivates the construction of  cubic-kilometer observatories. 
In order to meet the Gigaton mass scale  and the need to be in underground, natural transparent materials are used as detector material.
The projects of this scale are: IceCube \cite{IceCube} installed in the 2800~m thick glacial ice at the South Pole and in data taking stage,  KM3NeT \cite{km3net} \cite{P}  under design for being operated in ocean and the Gigaton Volume Detector in Lake Baikal (Baikal GVD) \cite{baikal} also under design and test.
As AMANDA in 12 years (1997-2009) of operation  and data analysis  opened the way  for  IceCube, Antares \cite{GA} installed in the Mediterranean provides important inputs for KM3NeT as well as, together with Baikal, a look in neutrinos to the Southern Hemisphere. \\
\noindent
Neutrino telescopes are versatile detectors. In addition to the physics goals summarized above, IceCube has the capability to detect neutrino bursts from nearby supernovae by exploiting the low photomultiplier noise in the antarctic ice (on average 280~Hz for IceCube) \cite{SN_ICRC} \cite{R. Maruyama}, to measure the primary composition of cosmic rays by analyzing events seen in coincidence by the air shower array IceTop \cite{icetop} and the deep strings of IceCube \cite{CR_ICRC} and to explore energies below 100~GeV with the use of the sub-detector DeepCore \cite{J. Koskinen} \cite{DC_ICRC}. As example, we anticipate here the recent observation of an anisotropy in cosmic rays in the southern hemisphere by IceCube \cite{anisotropy}.\\
\noindent
%how does a NT work? fov ?
A neutrino telescope detects HE neutrinos by observing the Cherenkov radiation from charged particles produced by neutrino interactions.
These secondary particles depend on the nature of the interaction (neutral or charge current) as well as on the energy of the incoming neutrinos.
The sources of background for the identification of extra-terrestrial neutrinos are muons and neutrinos from the decays of particles produced through the interaction of cosmic rays in the atmosphere.
The different event topologies used in order to separate signal from the atmospheric background are schematically described in Fig.~\ref{fig:signatures}. They can be separated in:\\
{\bf - Upwards through-going muon}: penetrating atmospheric muons are on average stopped after few kilometers of matter. As a consequence the observation of an upwards track implies that a  $\nu_\mu$ interacted via CC interaction in the vicinity of the detector. The field of view (fov) of a neutrino telescope based on upwards tracks is then restricted to the hemisphere opposite the geographical position of the detector. \\
{\bf - Very high energy downwards muon}: the atmospheric background rises steeply with the angle above the horizon and it is dominated by bundles of multiple muons.  Cherenkov emission muon bundles has a signature similar to a single very high energy muon track induced by a neutrino. At very high energies, the rate of bundles of muons is decreasing faster then a possible ultra high energy neutrino signal. The bundle background topology can be rejected efficiently on the base of energy observables \cite{RobertL}. This approach opens at energies above the PeV  the fov of a neutrino telescope to the hemisphere above the detector. \\
{\bf - Contained hadronic shower}: in NC interaction of $\nu_\mu$ and CC of $\nu_e$, $\nu_\tau$, the neutrino transfers a fraction of its energy to a nuclear target, producing an electromagnetic or hadronic particle shower. The main background for neutrino-induced cascade searches comes from the stochastic energy losses suffered by cosmic ray muons as they pass through the telescope \cite{Michelangelo}.  In order to discriminate these showers from stochastic energy losses, the shower has to be contained in a pre-defined fiducial volume. With this signature, a neutrino telescope is in principle sensitive to the entire sky.\\
{\bf- Contained and semi-contained muon}:  once a neutrino telescope like IceCube approaches the cubic-kilometer scale, a core or fiducial volume (FV)  nearly free from atmospheric muons can be identified. The principle is that atmospheric muons will be vetoed by the most external part of the detector itself (veto volume= VV).  Contained events will be tracks that do not leave any sign in the VV and "start" in the FV.  The probability that a starting track is induced by a neutrino is then  determined via MonteCarlo simulation.\\
More specific signatures dedicated to slow monopole or super-symmetric particles are not included in this discussion.

\section{IceCube}
\label{IceCubeSection}
\begin{center}
\begin{figure*}[ht]
\centering
\includegraphics[scale=0.38]{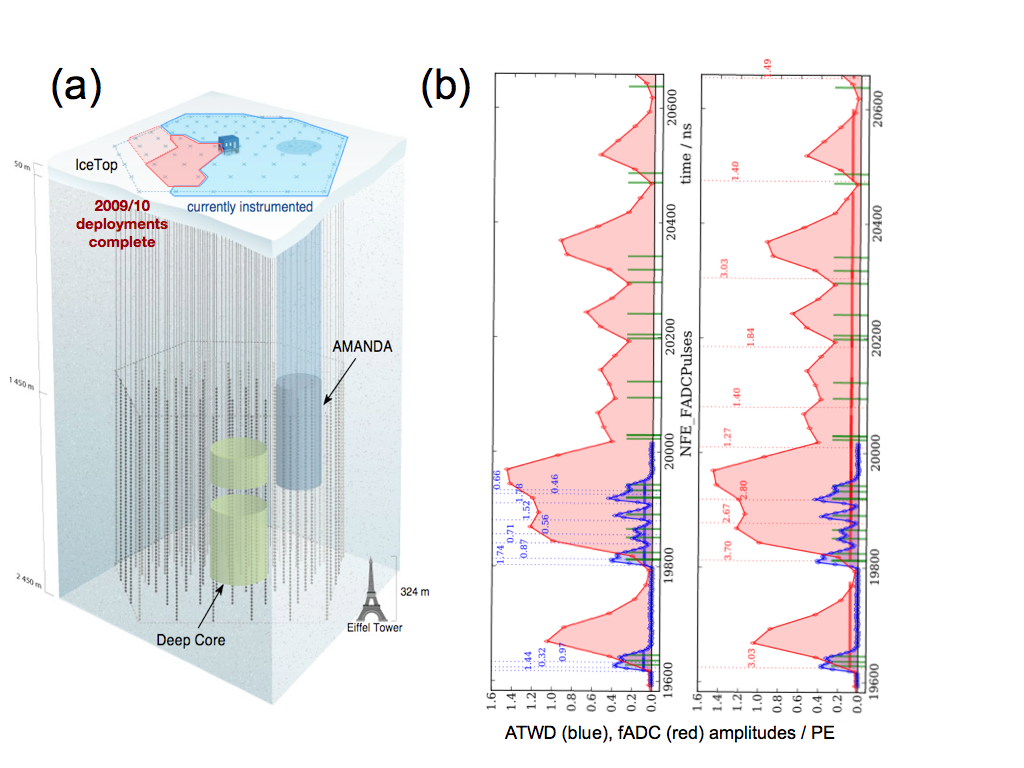}
\caption{(a) IceCube conceptual scheme. (b) Example of waveform sampling and feature extraction.}
\label{IceCube-DOM-W}
\end{figure*}
\end{center}

\noindent
%status
IceCube \cite{IceCube} \cite{IceCube2}, the largest neutrino telescope in history is now a reality.
The 2800~m thick glacial ice sheet at the South Pole is used by IceCube  as a Cherenkov radiator for charged particles.
The Cherenkov light produced by the collision of cosmic neutrinos with subatomic particles in the ice or nearby rock  is detected by an embedded array of Digital Optical Modules (DOMs). Each DOM  incorporates a 10 inches diameter R7081-02 photomultiplier tube (PMT) made by Hamamatsu Photonics. The finished array will consist of 5160 DOMs at depths of 1450-2450~m attached to 86 vertical strings. Moreover, 80 IceTop surface stations provide a unique air shower detector for the study of primary cosmic rays. The layout of the array is shown in Fig.~\ref{IceCube-DOM-W}(a).
At present, 79 strings are installed and operational, with the remaining 7 strings to be deployed in the coming austral summer. A summary of the deployment seasons and IceCube configurations is reported in Tab.~\ref{history}. Also shown in Fig.~\ref{IceCube-DOM-W}(a) is the AMANDA telescope, which operated as an independent instrument from 2000-2006 and was integrated into IceCube from 2007-2008. AMANDA consisted of 687 optical modules on 19 strings, with the bulk of the detector at depths of 1500 to 2000 m. 
%The AMANDA optical modules were simpler than their IceCube counterparts, with considerably smaller dynamic range and less ability to resolve individual photoelectrons. 
In 2009, AMANDA was decommissioned and the first string of DeepCore  \cite{J. Koskinen} was deployed. \\
\noindent
A $\nu_\mu$ with energy greater then 100~GeV interacting in or around IceCube instrumented volume creates a muon that traverses kilometers of ice and generates Cherenkov light along its path. Above 1~TeV, the muon loses energy stochastically to produce multiple showers of secondary particles, resulting in an overall light yield proportional to the muon energy \cite{muon1} \cite{muon2}.  For DOMs close to the incoming track, most photons arrive in a pulse less than 50~ns wide, in which the earliest photons have traveled straight from the muon track without scattering. Significant scattering  (\cite{IcePaper}) lengthens the  light pulses with distance; the light pulses reach 1~$\mu$s (FWHM) for DOMs 160~m away from a muon track. 
%Depending on primary energy and distance from the track, each PMT can see single photons or pulses ranging up to thousands of photons.
The elements that make IceCube function are: DOM's time response, optical sensitivity, time dispersion, optical attenuation introduced by the ice.
PMTs' waveforms are extracted and correlated in order to reconstruct the incoming direction and energy of the muon.
%Each maximum likelihood fit is based on the complete pattern of light amplitude and timing seen by the DOMs.  
Design studies for IceCube physics goals \cite{IceCube2} have shown that sufficient reconstruction quality is achieved for a PMT timing resolution of 5~ns, low-temperature noise rate below 500~Hz, and effective dynamic range of 200 photoelectrons per 15~ns. The description of the characterization of IceCube PMTs is reported in \cite{PMTcal}.\\
%IceTop hardware
IceTop uses DOMs identical to those in the deep ice. Here the signals arise from muons, electrons and gamma rays in cosmic ray air showers \cite{icetop}. These particles deposit energy in the ice tanks housing the DOMs, resulting in light pulses up to several hundred nanoseconds long. The arrival times and amplitudes in the surface array are then used to reconstruct the shower core position, direction, and energy. An overall timing resolution of 10~ns provides pointing accuracy of about one degree. The PMT pulses range from single photoelectrons at the periphery of showers to $10^5$ photoelectrons for a 1~EeV shower that strikes within the array. To achieve the implied dynamic range, each tank contains two DOMs operating at gains differing by a factor 50.\\
%DAQ
The analog signal produced by the PMT is digitalized in the DOM by a custom made Analog Transient Waveform Digitizer (ATWD) and an fADC  (\cite{DAQ-S}). The data from a single trigger  consists of at least one ATWD waveform and one fADC waveform, plus a time stamp and the local coincidence signals from the adjacent DOMs.
%IceCube software / waveform feature extraction
The number and time distribution of photons registered in calibrated waveforms are determined via a feature extractor software \cite{Marius}. 
Pulse shape functions from the PMT and the fADC have to be deconvolved by efficient and robust algorithms. An example of correctly interpreted waveform is reported in Fig.~\ref{IceCube-DOM-W}(b). The IceCube software is built within a highly modular framework called IceTray \cite{IceTray}. Software environments for specialized tasks such as online-filtering, simulation (IceSim), reconstruction (IceRec), or analysis are bundled to form different meta-projects. The simulation needs of IceCube are very demanding; for this reason a software package in Python has been written in order to manage, run, control and monitor the generation of the IceCube detector simulation data and related filtering and reconstruction analyses (IceProd \cite{IceProd}). Simulation production is distributed among university computer clusters in USA, Germany and Sweden.

\begin{center}
\begin{table}[htb]
%{\small
%\hfill{}
\scalebox{0.72}{%
\begin{tabular}{ c | c | c | c | c |c}
  Nr.  & Deploy.  & Physics Run & AMANDA & DeepCore & Nr  $\nu_{atm}$\\
  Strings   & Season           &  (Start-Stop)   &  Status     &  &\\
  \hline \hline
  1 & 04-05 & -- & Running & -- & --\\
  \hline
  9 & 05-06 &  Start: 2006-06 & Running & -- & 234 \cite{JP} \\
     &              & Stop: 2006-11                                              &                  &   & (based on \\
     &               & Total: 137.4~d                &                   &     & 6 months) \\
  \hline
  22 & 06-07 & Start: 2007-02-16 & Running  & Design  & 5,114 \cite{IC22}\\
        &            & Stop: 2008-04-5 &   (integrated  & Study \cite{Olaf} & \\    
        &            & Total: 275.70~d & mode)   & \\  
  \hline                           
  40 & 07-08 & Start: 2008-04-05 & Running  & 6 strings  &   14,121  \cite{JD}\\
          &            & Stop: 2009-05-20   &   (integrated & approved & upwards \\\    
        &            & Total: 375.5~d &   mode)  & &   \\  
  \hline
  59 & 08-09 & Start: 2009-05-20  & Decommis-& 1 string &  $\sim$25,000 up \\
            &            & Stop: 2010-05-30  &   sioned  & deployed  & upwards \\    
        &            & Totoal: $\sim$350~d &    & \\  
  \hline
  79 & 09-10 & Start: 2010-05-31  & --  & 6 strings &  --\\
              &            & Still running  &    & deployed &\\    
   \hline
  86 & 10-11  & --  & -- & --& --\\
\end{tabular}}
\caption{Summary table of IceCube deployment, physics runs, Deep Core status and atmospheric neutrinos detected.}
\label{history}
\end{table}
\end{center}

\vspace*{-18mm}
\subsection{IceCube preliminare performances}
\begin{figure}[ht]
\centering
\includegraphics[scale=0.27]{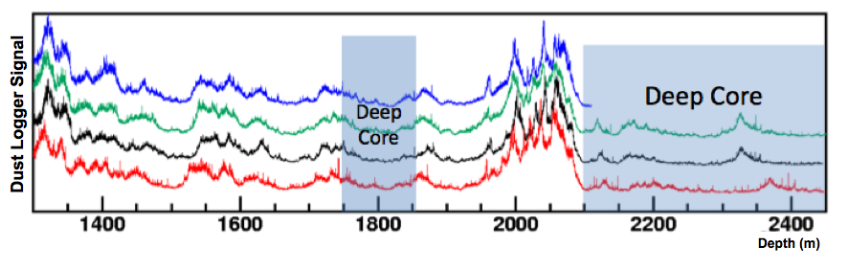}
\caption{Dust concentration in the South Pole ice versus depth. }
\label{dust}
\end{figure}

\noindent
After more then four years of operation of the IceCube array (see Tab.~\ref{history} for IceCube configurations), preliminary performances can be assessed.\\
{\bf - Detection Material: the Ice}\\
\noindent
The South Pole ice is one of the purest and most transparent natural material on Earth. 
The amount of dust from volcanic ash in the ice is between 0.05-2~ppm, depending on the depth. 
If we assume that the concentration of U/Th/K is the same as in the EarthÕs crust  then we expect concentrations of U (0.1-8~pg/g), Th (5-200~pg/g) and K (0.75-30~ppb) \cite{ashes}. The deepest ice at the geographical South Pole ($\sim$2500~m) is found to be in the lower range of these numbers. In situ measurements of drilling water of IceCube holes support this trend.
Moreover, the ice is completely inert, dry, with an extremely low ambient noise and with no significant variation on time scales of years. 
The IceCube Collaboration has invested since the AMANDA time a considerable effort in order to map the optical properties, temperature dependences and dust concentration \cite{dustlog} of the ice. Pulsed and continuous light sources at various wavelength and depth are regularly used in order to determine physics optical properties \cite{IcePaper}. 
Thanks to the constant calibration effort, the IceCube Collaboration has been able to transform a remote and harsh environment into a usable radiation detection material.\\
\noindent
{\bf - Detection Sensors: the DOMs}\\
\noindent
More then 5000 DOMs are operating in ice (underground array  or IceTop tanks). An overall failure rate of $\sim$1\% is observed: about 65 DOMs did not work after deployment and about 15 have broken during data taking \cite{Mark}. 
The characteristic noise of the DOM is very low: the average rate per DOM (200~$\mu s$ artificial delay) corresponds to $284\pm 26.2$~ Hz. The  DeepCore DOMs are equipped with high quantum efficiency PMTs. The average dark noise is slightly higher and corresponds to $358.9\pm 36.0$~Hz. 
Monthly calibrations (DOMCal) are performed in order to monitor DOM-by-DOM calibration constants like PMT gain as a function of HV, analog front end gains and offsets, discriminator thresholds, and digitizer sampling speed. These calibrations  show minimal changes in the DOM (mainly due to seasonal variations). The overall stability of the DOMs guarantee a uniform and quiet data taking along the entire physics run. In addition, we note that no specific failure has been observed in cables or connectors confirming that operation in ice are extremely reliable and stable.\\
\noindent
{\bf - Data Management}\\
\noindent
The overall data taking live time is on average 98\% per physics run. Part of the data taking is devoted to calibration and maintenance of the detector. 
As a consequence, the live time effectively used for data analysis is on average 93\%.
Each single isolated hit is acquired from the entire detector. The global trigger rate for IC79 configuration is  2~kHz.
Control operations on various subsystems and the historical state of the detector are performed via a dedicated experiment control system called IceCube Live (I3Live) \cite{i3live}.
The volume of data produced by the data acquisition system far exceeds the limited bandwidth available in the  satellite allowance. 
Instead of taping the entire data sample, an online filtering system is used to apply a set of first-level event selections to the collected data. 
A dedicated effort for the maintenance of the online system succeeds to provide high quality data for physics analysis.\\
\noindent
{\bf - Performances}\\
\noindent
In order to understand and monitor the performances of the IceCube detector a series of low level quantities are used. Typical variables are: the number of hits DOMs (NChannel), the number of hits that each DOM receives in a given run (DOM occupancy),  the center of gravity of all PMT hits of an event  weighted by the charge per PMT (COG), the number of direct hits etc. Moreover, reconstructed variables and relative quality parameters are used like the zenith and azimuth distribution of the events at different purity level and the energy distribution based on different reconstruction algorithms. Extensive comparisons with MonteCarlo simulation reveals the status of the understanding. Overall the agreement between experimental and simulated data is satisfactory. Discrepancies are observed in particular in the occupancy plots versus the vertical direction. These discrepancies are concentrated in the bottom part of the detector where the ice properties are extremely good as revealed by in-situ measurements (see Fig.~\ref{dust}). The detailed optical properties of the deep ice are still under optimization. In situ light sources (LED) are used in order to extract the optical properties of the ice which are then input  in the simulation chain. Possible tilt in the deep ice layers as well as wavelength dependencies are still under study.

%\begin{center}
%\begin{wrapfigure}{r}{0.5\textwidth}
%\hspace*{-40mm}
%\end{wrapfigure}
%\end{center}

\noindent
A complementary strategy for the study of IceCube performances uses high level analysis like the so called shadow of the moon  \cite{moon}: the absolute pointing accuracy and the angular resolution of the telescope is experimentally tested via the study of the deficit of cosmic rays from the direction of the Moon.
 IceCube has observed the shadow of the Moon as a 5.0~$\sigma$ deviation from event counts in nearby regions, using data from 8 of the total 13 lunar months in the data taking period with the 40-string detector setup. From this, we can conclude that IceCube has no systematic pointing error larger than the search bin, 1.25$^o$ \cite{moon}.  \\
An additional probe of the correct functioning of the detector comes from the recent measurement of  a small but very interesting anisotropy of cosmic ray arrival directions \cite{anisotropy}.
The data used for the first anisotropy measure are the one from IC22 and included 4.3 billion muons produced by down-going cosmic ray interactions in the atmosphere; these events were reconstructed with a median angular resolution of 3$^o$ and a median energy of  about 20~TeV. Their arrival direction distribution exhibits an anisotropy in right ascension with a first harmonic amplitude of $(6.4 \pm 0.2 stat. \pm 0.8 syst.) \cdot 10^{-4}$. This measure is unique in the Southern Hemisphere but matches nicely with similar measurements in the Northern Hemisphere. Such a precise measurement could not have been performed in case of disfunction of the telescope.

\vspace*{-3mm}
\section{IceCube results}
\noindent
IceCube is a discovery instrument. 
%The goal of IceCube analysis is to reject or accept the so called null hypothesis H0 (i.e. the background only hypothesis) in an unbiased and correct (i.e. not false) way.
In order to avoid practitioner biases, IceCube uses blind analysis. 
%In searches like the one for an astrophysical diffuse flux, part of the data are kept completely closed and the cuts are optimized on a part of the data (burn sample); in analysis like the point source search,  analysis variables are scrambled (for example the azimuth).
A fairly long list of searches (multiple comparisons or multiple tests) are typically realized on IceCube data samples: this is due to the fact that  it is not known a priori which category of source (AGNs, GRBs, galactic sources etc), region of the sky (galactic or extragalactic) or mechanism (cosmic ray interaction, dark matter annihilation, exotic mechanisms) will produce a visible extraterrestrial neutrino signal. 
The larger the number of IceCube tests (or searches) becomes,  higher is the chance of a false discovery claim. In order to avoid false discoveries, the IceCube collaboration has the policy to re-correct final probabilities for a trial factor and  fix a uniform discovery threshold for all the tests (5~$\sigma$).\\
\noindent
In this paper, we choose to not prioritize few single results but instead to summarize nearly the whole family of IceCube hypothesis tested and related references. The aim is to show the richness of the IceCube program and the status.
The  {\bf IceCube 22-strings data sample} corresponds to a total live time  of 275.70~days.  Most of the tests performed on the IceCube 22-strings are summarized in Tab.~\ref{IC22}.
A total of 27 hypothesis have been tested in a blinded fashion on the 22-strings sample. Among these searches, four have presented a p-value of the order of 1\%. Then, following the IceCube policy, no discovery has been claimed and 90\%CL upper limits are calculated.
The  {\bf IceCube 40-strings data sample} corresponds to a total live time of 375.5~days. The analysis of this sample is presently (October 2010) not jet complete. Until now,
a total of 25 hypothesis have been tested in a blinded fashion on the 40-strings sample. Preliminary results from some of the performed searches are reported in Tab.~\ref{IC40} with related references.

\section{Final Remarks}
\noindent
The visionary dream of a 1~km$^3$ neutrino telescope at the South Pole turned  into a firm reality with the installation and operation of IceCube.  
A total of 79 strings, 73 IceTop tanks, more then 5000 DOMs operating in stable mode provide the clear
evidence that the IceCube technology is mature and successful. 
Data analysis of half of the full detector shows stable performances of the telescope and solidity of the hardware. 
Constraints have been placed on the most optimistic model predictions.
The IceCube Collaboration is very close to meeting the challenge goal of 86 strings / 80 IceTop tanks. 
Optimal conditions for a discovery are provided. 
The IceCube Collaboration is probing nature's mysteries in an unprecedented way. 
Do you think we will stop here?

%% The Appendices part is started with the command \appendix;
%% appendix sections are then done as normal sections
%% \appendix

%% \section{}
%% \label{}

%% References
%%
%% Following citation commands can be used in the body text:
%% Usage of \cite is as follows:
%%   \cite{key}         ==>>  [#]
%%   \cite[chap. 2]{key} ==>> [#, chap. 2]
%%

%% References with BibTeX database:

%\bibliographystyle{elsarticle-num}
%\bibliography{<your-bib-database>}

%% Authors are advised to use a BibTeX database file for their reference list.
%% The provided style file elsarticle-num.bst formats references in the required Procedia style

%% For references without a BibTeX database:

%\pagebreak

\begin{center}
\begin{table*}[ht]
%{\small
%\hfill{}
\scalebox{0.87}{
\begin{tabular}{ l | c | c }
   Physics Analysis  &  Results & Reference \\
  \hline \hline
  Steady Point Source Search-   & Unbinned likelihood point source search & \cite{IC22}\\
   Northern Hemisphere                             & No evidence for a point source.   & \\
                   & Average upper limit  $E^2 \Phi_{\nu_\mu} < 1.4 \cdot 10^{-11} TeV cm^{-2} s^{-1}$ [3~TeV - 3~PeV]    &\\   
   \hline 
   Steady Point Source Search-      &  Field of view opened to region above the horizon. &\\  
   Above the horizon	                & 28 sources tested, for example:  & \cite{RobertL}\\
                                                        & Galactic center limit $E^2 \Phi_{\nu_\mu} < 2.4 \cdot 10^{-10} TeV cm^{-2} s^{-1}$ [1.3 PeV - 890~PeV]    &\\
                                                        & Centaurus A limit $E^2 \Phi_{\nu_\mu} < 5.6 \cdot 10^{-10} TeV cm^{-2} s^{-1}$ [1.3 PeV - 890~PeV]    &\\  
   \hline
   Steady Point Source Search & Search optimized for soft spectra galactic sources & \\
   For lower energies (with AMANDA) & No evidence for a point source. & \cite{Yolanda}\\
                                                                    & $E^{-2.4}, E{_cutoff} = 50~TeV$  & \\
							& $1.2 \cdot 10^{-10} TeV^{-1}cm^{-2}s^{-1}$ &\\
   \hline
   Dark Matter:  & Neutralino masses tested: 250 - 5000~GeV & \cite{WIMPs}\\
  Neutralino & most stringent limits on neutralino annihilations in the Sun & \\
   &  most stringent limits on the spin-dependent WIMP-proton & \\
   &cross-section for neutralino masses above 250 GeV & \\ 
  Lightest Kaluza-Klein (LKP) &  LKP masses tested: 250 - 3000~GeV &   \cite{KK}\\ 
   & most stringent  limits on the LKP-proton cross-sections&\\
   \hline
   GRBs Search & 41 gamma- ray bursts (GRBs) in the northern sky & \cite{GRB22}\\   
                            & upper limit on the fluence from:  & \\   
                            & prompt phase $3.7 \cdot 10^{-3}$ erg cm$^{-2} $ [72~TeV - 6.5~PeV]  & \\
                            & precursor phase $ 2.3 \cdot 10^{-3}$ erg cm$^{-2}$ [2.2~TeV - 55~TeV] & \\
   \hline
   Extra-terrestrial neutrino-induced  & $E^2\phi_{\nu_e+\nu_\mu+\nu_\tau}  <  3.6 \cdot 10^{-10}~TeV  s^{-1}  sr^{-1}$& \cite{jo} \\
   cascades search & [24~TeV - 6.6~PeV] & \\
   \hline
   Extremely high energy (EHE) & No evidence of an astrophysical flux.& \cite{EHE}\\
   cosmogenic neutrinos  search & Upper limit $E^2\phi_{\nu_e+\nu_\mu+\nu_\tau} < 5.6 \cdot 10^{-10}$~TeV cm$^{-2} s^{-1}sr^{-1} $& \\
   &  [$10^{7.5} < E_\nu < 10^{10.6}$~ GeV] & \\
  
  \end{tabular}
 }
 
%\hfill{}
\caption{Partial list of IceCube searches and related references realized on the 22 strings and AMANDA sample. A total of 27 hypothesis tests realized in a blinded fashion have been performed. All upper limits quoted are at 90\% confidence level.}
\label{IC22}
\end{table*}
\end{center}

\begin{center}
\begin{table*}[ht]
%{\small
%\hfill{}
\scalebox{0.87}{%
\begin{tabular}{ l | c | c }  
    Physics Analysis  &  Results & Reference \\      
  \hline \hline                          
  Steady Point Source Search - &  First all-sky point source search & \cite{JD}\\
   All-sky  & Upper limit northern sky $E^2 \Phi_{\nu_\mu} \sim 2-200 \cdot 10^{-12} TeV cm^{-2} s^{-1}$ [TeV - PeV]    &  \\
                                               & Upper limit southern sky $E^2 \Phi_{\nu_\mu} \sim 3-700 \cdot 10^{-12} TeV cm^{-2} s^{-1}$ [$>$~PeV]  & \\
        &  No evidence of a point source. & \\
 \hline
  Transient Point Source Search & Blazar multi-wavelength flares search & \\
   					            & Microquasar Periodicity test & \cite{levent} \\
					            & Model independent flare search & \\
					            & No evidence of a point source & \\
   \hline
   GRBs Search & 117 gamma- ray bursts (GRBs) in the northern sky & \cite{kevin}\\   
                             &  First analysis sensitive to the flux predicted by fireball phenomenology, & \\
                            & upper limit on the fluence: & \\   
                            &  $1.1 \cdot 10^{-3}$ erg cm$^{-2} $ [37~TeV - 2.4~PeV]  & \\ 
                            & Guetta et al. flux \cite{guetta} excluded at 90\% confidence in the region $\pm 2248$ seconds  & \\
   \hline   
   \hline
   Extra-terrestrial diffuse neutrino flux  & $E^2\phi_{\nu_\mu}  <   8.9 \cdot 10^{-12}~TeV  s^{-1}$  cm$^{-2} sr^{-1}$& \cite{tomdiffuse} \\
   search & valid in the range [34.7~TeV - 6.9~PeV] & \\
                &  First upper limit below the Waxman Bachall  bound & \\
                &   Models from Becker et al \cite{agn1}, Stecker et al. \cite{agn2} excluded at 90\% confidence & \\
   \hline
   Extremely high energy (EHE) & No evidence for an astrophysical flux.& \cite{aya40}\\
   cosmogenic neutrinos  search & Upper limit $E^2\phi_{\nu_e+\nu_\mu+\nu_\tau} <  4.23 \cdot 10^{-11}$~TeV cm$^{-2} s^{-1}sr^{-1} $& \\
   &  [$10^{6.3} < E_\nu < 10^{9.8}$~ GeV] &  \\
   
  \end{tabular}
  }
%\hfill{}
\caption{Partial list of IceCube searches realized on the 40 strings and AMANDA sample. Results reported are still preliminary. A total of 25 hypothesis tests have been realized in a blinded fashion. All upper limits quoted are at 90\% confidence level.}
\label{IC40}
\end{table*}
\end{center}

%\end{linenumbers}

\end{document}